\documentclass{emulateapj}
\usepackage{float}
\usepackage[utf8]{inputenc}
\usepackage{xcolor}
\usepackage{hyperref} 
\hypersetup{backref=true, pagebackref=true, hyperindex=true, colorlinks=true, breaklinks=true, urlcolor= blue,                
    linkcolor= blue, bookmarks=true, bookmarksopen=false, filecolor=blue, citecolor=blue, linkbordercolor=blue
}
\AtBeginDocument{\hypersetup{pdfborder={0 0 1}}}

\shorttitle{The Quasar Ensemble Variability with DECaLS and SDSS}
\shortauthors{Li et al.}

\begin{document}

\title{THE ENSEMBLE PHOTOMETRIC VARIABILITY OF OVER $10^5$ QUASARS IN THE DARK ENERGY CAMERA LEGACY SURVEY AND THE SLOAN DIGITAL SKY SURVEY}

\author{Zefeng Li\altaffilmark{1}, Ian D. McGreer\altaffilmark{2}, Xue-Bing Wu\altaffilmark{1, 3}, Xiaohui Fan\altaffilmark{2, 3}, and Qian Yang\altaffilmark{1, 3}}

\altaffiltext{1}{Department of Astronomy, School of Physics, Peking University, Beijing 100871, China; zefeng.li@pku.edu.cn}
\altaffiltext{2}{Steward Observatory, University of Arizona, 933 North Cherry Avenue, Tucson, AZ 85721, USA}
\altaffiltext{3}{Kavli Institute for Astronomy and Astrophysics, Peking University, Beijing 100871, China}

\begin{abstract}
We present the ensemble variability analysis results of quasars using the Dark Energy Camera Legacy Survey (DECaLS) and the Sloan Digital Sky Survey (SDSS) quasar catalogs. Our dataset includes 119,305 quasars with redshifts up to 4.89. Combining the two datasets provides a 15-year baseline and permits analysis of the long timescale variability. Adopting a power-law form for the variability structure function, $V=A(t/1yr)^{\gamma}$, we use the multi-dimensional parametric fitting to explore the relationships between the quasar variability amplitude and a wide variety of quasar properties, including redshift (positive), bolometric luminosity (negative), rest-frame wavelength (negative), and black hole mass (uncertain). We also find that $\gamma$ can be also expressed as a function of redshift (negative), bolometric luminosity (positive), rest-frame wavelength (positive), and black hole mass (positive). Tests of the fitting significance with the bootstrap method show that, even with such a large quasar sample, some correlations are marginally significant. The typical value of $\gamma$ for the entire dataset is $\gtrsim 0.25$, consistent with the results in previous studies on both the quasar ensemble variability and the structure function. A significantly negative correlation between the variability amplitude and the Eddington ratio is found, which may be explained as an effect of accretion disk instability.

\end{abstract}

\keywords{galaxies: active – galaxies: nuclei – quasars: general – techniques: photometric}

\section{Introduction}
Quasars were observed to be variable soon after their discovery \citep[]{Matthews63}.  Variations were found in the optical and other wavebands, and the variation timescales range from hours to years \citep[e.g.][]{de03, McHardy06, Sesar06, Wilhite08, Kasliwal15}. The typical variability amplitude is several tenths of one magnitude in the optical band \citep{Smith95, Giveon99, Collier01, Mac10}. Variability is a useful method to select quasars \citep{van73, ivezic04, Rengstorf06, Schmidt10, Morganson14}. Recently, many studies have presented the results of variabilities of multi-band optical photometric surveys, such as SDSS \citep[Sloan Digital Sky Survey,][]{Vanden Berk04}, OGLE \citep{Kozlowski10}, the Palomar-QUEST Survey \citep{Bauer09}, and the SDSS Stripe 82 \citep{Sesar07, Mac10, Zuo12}. Comparisons between wide-area surveys with long time separations provide insight into quasar variability properties on long timescales, such as SDSS-POSS \citep[Palomar Observatory Sky Survey,][]{Mac12} and SDSS-PS1 \citep[Pan-STARRS1,][]{Morganson14}.

Although the physics of the quasar variability remains unclear, several possible explanations have been proposed, including accretion disk instabilities and variations in the accretion rate \citep[e.g.][]{Rees84, Kato96, Aretxaga97, Kawaguchi98, Czerny08}, nuclear supernovae \citep[e.g.][]{Terlevich92, Cid97}, star-star collisions \citep{Courvoisier96, Tor00}, and gravitational microlensing by compact foreground objects \citep[e.g.][]{Chang79, Hawkins93, Lewis96, Zackrisson03}.  Recently, a damped random walk model was proposed to explain quasar lightcurves \citep[e.g.][]{Kelly09, Mac10, Mac12}, and it was interpreted to be consistent with the expectation from thermal fluctuations \citep[e.g.][]{Dexter10}.

One approach to characterizing quasar variability is to perform an ensemble study. The average variability as a function of time is determined for a large number of quasars, although individual quasars have disparate light curves \citep[e.g.][]{Mac10}. These differences are smeared or erased when averaging the quasar properties. Ensemble studies allow quasar variability to be studied for large quasar samples with highly sparse lightcurves, and are well suited for comparing two independent surveys that provide only approximately two photometric data points per quasar. A well-established approach to quantifying quasar variability from ensemble studies is to use the structure function, which measures the amplitude of variability as a function of time lag.

\cite{Angione72} reported an anti-correlation between the quasar variability amplitude and luminosity and it was soon confirmed by a large number of studies \citep{Uomoto76, Pica83, Lloyd84, OBrien88, Hook94, Trevese94, Cid96, Cristiani96, Paltani97, Giveon99, Garcia99, Hawkins00, Webb00, Vanden Berk04, Wilhite08, Kelly09, Mac10, Zuo12, Morganson14, Caplar16}. A similar anti-correlation between the variability amplitude and rest-frame wavelength was also elucidated by \cite{Vanden Berk04}, \cite{Mac10}, \cite{Zuo12}, and \cite{Morganson14}. However, previous studies did not reach a consensus as to black hole mass dependence; correlations were found to be positive \citep{Wold07, Wilhite08, Ai10}, negative \citep{Kelly09}, or uncertain \citep{Zuo12, Caplar16}.

In previous studies, the sample number varied from nearly 100 \citep{Wold07} to nearly 20,000 \citep{Bauer09}. The SDSS has greatly increased the number of quasars with spectroscopic redshifts. \citep{Schneider10, Paris17}. Along with other surveys, SDSS-PS1 ensemble variability has been studied \citep{Morganson14}, including 105,783 identified quasars and a wide time lag range from 0.01 to 10 years in the rest-frame. The Dark Energy Camera Legacy Survey (DECaLS) is a new survey for studying quasar statistical variability. Thanks to SDSS-DECaLS overlap regions, our quasar population reaches nearly 120,000, the time lag reaches 10 years, and the highest redshift reaches 4.89. Our dataset covers a wide range of parameters for the quasar properties including redshift and bolometric luminosity, which, along with the DECaLS deep field survey, will provide more nearly complete results.

In this paper, we will discuss the main features of a specific quasar dataset established by combining SDSS and DECaLS. In Sections \ref{Section_2}, we describe the SDSS-DECaLS dataset, the magnitude calibration, and the photometric noise estimation. In Section \ref{Section_3}, the structure function is discussed. In Sections \ref{Section_4} and \ref{Section_5}, we investigate the fitting parameters of the variability dependency on the quasar properties. Finally, in Sections \ref{Section_6} and \ref{Section_7} we discuss the results, as well as the comparisons with several previous studies and provide a summary. Photometric data from SDSS are in the SDSS photometric system \citep{Lupton99}, which is almost identical to the AB system. Since DECaLS magnitude is in the AB system \citep{Oke83}, we use the AB system for both SDSS and DECaLS throughout this paper.

\section{DATASET}
\label{Section_2}
\subsection{The SDSS-DECaLS Dataset}
\label{Section_2.1}
The SDSS project covers a sky area of $\sim$14,000 $\mathrm{deg^2}$, mainly in the northern Galactic cap, including five broad bands \citep[$u, g, r, i,$ and $z$,][]{Doi10}. SDSS Data Release 12 quasar catalog (DR12Q) includes $\sim$300,000 quasars observed in the imaging survey \citep[][]{Paris17}. From the SDSS DR12 quasar catalog, we get flux, magnitude, inverse variance of flux (IVAR), and photometric Modified Julian Date (MJD) data. We also include bolometric luminosity, black hole mass and the Eddington ratio of each quasar by adding the SDSS DR12Q black hole mass catalog \citep[][]{Kozlowski16}. However, DR12Q only contains quasars observed in SDSS-III. We add Data Release 7 quasars (DR7Q) \citep[][]{Schneider10, Shen11} to fill the gap in redshift from $\sim$1.0 to $\sim$2.0 and to increase the number of bright quasars in our study.

The ongoing Legacy Survey is producing a model catalog of the sky from a set of both optical and infrared imaging data, aiming to comprise 14,000 $\mathrm{deg^2}$ of extragalactic sky visible from the northern hemisphere. The sky coverage is approximately bounded in the range of $-18^{\circ}$ $\le \delta \le$ +84$^{\circ}$ in celestial coordinates and $|b| > +18^{\circ}$ in Galactic coordinates. The Dark Energy Camera Legacy Survey provides data in the equatorial region in the range of $\delta < +30^{\circ}$ with the Dark Energy Camera (DECam) on the Blanco Telescope, in three optical bands $(g_\mathrm{DECam}, r_\mathrm{DECam}$, and $z_\mathrm{DECam})$. Flux measurements are obtained with the Tractor \citep{Lang16}, which uses a model-fitting approach to obtain catalog measurements. The DECaLS Data Release 3 (DR3) catalog is a great improvement over the previous DR2 catalog, covering 4,300 $\mathrm{deg^2}$ in $g_\mathrm{DECam}$ band, 4,600 $\mathrm{deg^2}$ in $r_\mathrm{DECam}$ band, and 8,100 $\mathrm{deg^2}$ in $z_\mathrm{DECam}$ band. Nearly 60\% of SDSS coverage is overlapped in DECaLS DR3, making this work on a large quasar population possible. For the magnitude limits, the median 5$\sigma$ point source depths for areas with 3 observations reach $g_\mathrm{DECam}=24.65, r_\mathrm{DECam}=23.61$, and $z_\mathrm{DECam}=22.84$. Excepting the MJD information, other content of SDSS are also shown in DECaLS DR3\footnotemark[4]. Temporarily DECaLS MJD information is not available in these files, so we adopt a method to get access as described in the Appendix.

\footnotetext[4]{http://portal.nersc.gov/project/cosmo/data/legacysurvey/dr3 \\ /external/survey-dr3-DR12Q.fits and http://portal.nersc.gov/project/cosmo/data/legacysurvey/dr3 \\ /external/survey-dr3-DR7Q.fits}

\begin{figure}
\centering\includegraphics[width=1.0\linewidth]{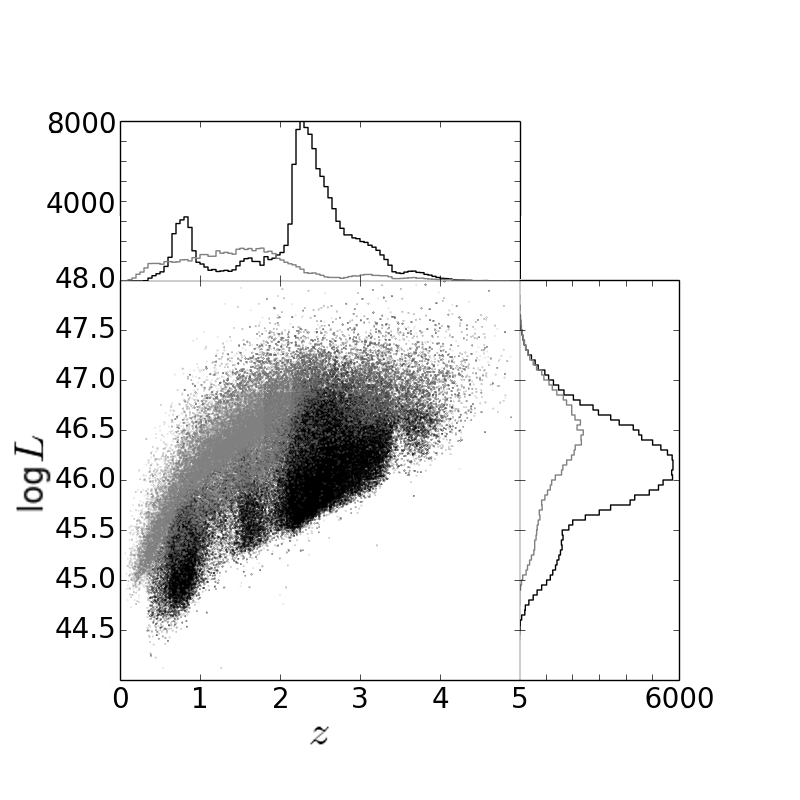}
\caption{Bolometric luminosity vs. redshift and the distributions of bolometric luminosities ($right$) and redshifts ($top$) for our quasar sample. The gray dots and histograms represent DR7 data while the black dots and histograms represent DR12 data. DR7Q fill in the redshift gap of DR12Q.}
\label{Figure_L_z_distribution}
\end{figure}

In this paper, we use the PSF magnitudes in both SDSS and DECaLS to make sure the results are accurate for point sources. We establish the dataset by concatenating DR12Q and DR7Q, both of which have the one-to-one matched counterparts in DECaLS DR3. Our quasar population consists of 119,305 quasars. Since the DECaLS $z_\mathrm{DECam}$ band covers a much larger area than the other two bands, a larger number of quasars have measurements in the $z_\mathrm{DECam}$ band, although the $g$- and $r$-band samples still exceed $\sim$50,000. The redshift range is from 0.06 to 4.89 (see Figure \ref{Figure_L_z_distribution}).

\subsection{The Magnitude Calibration between SDSS and DECaLS}
\label{Section_2.2}

Considering the different characteristics of the filters in the two surveys, the magnitude calibration is designed to calculate the transformation formulas and check the magnitude difference dependency on the color of the non-variable sources. We include SDSS standard stars \citep{ivezic07} and use the DECaLS file\footnotemark[5] for cross-matching, with the positional offset $<$1". In addition, we must eliminate those calibration stars that could be detected in one band in SDSS but not in the same band of DECam. Eventually 15,736 standard stars that are simultaneously detected in both DECaLS and SDSS are included.

\footnotetext[5]{http://portal.nersc.gov/project/cosmo/data/legacysurvey/dr3 \\ /survey-dr3-specObj-dr13.fits. A portion of the calibration stars \\ have spectra observed by SDSS.}
\footnotetext[6]{http://legacysurvey.org/dr3/description/}

\begin{figure}[t]
\centering\includegraphics[width=1.0\linewidth]{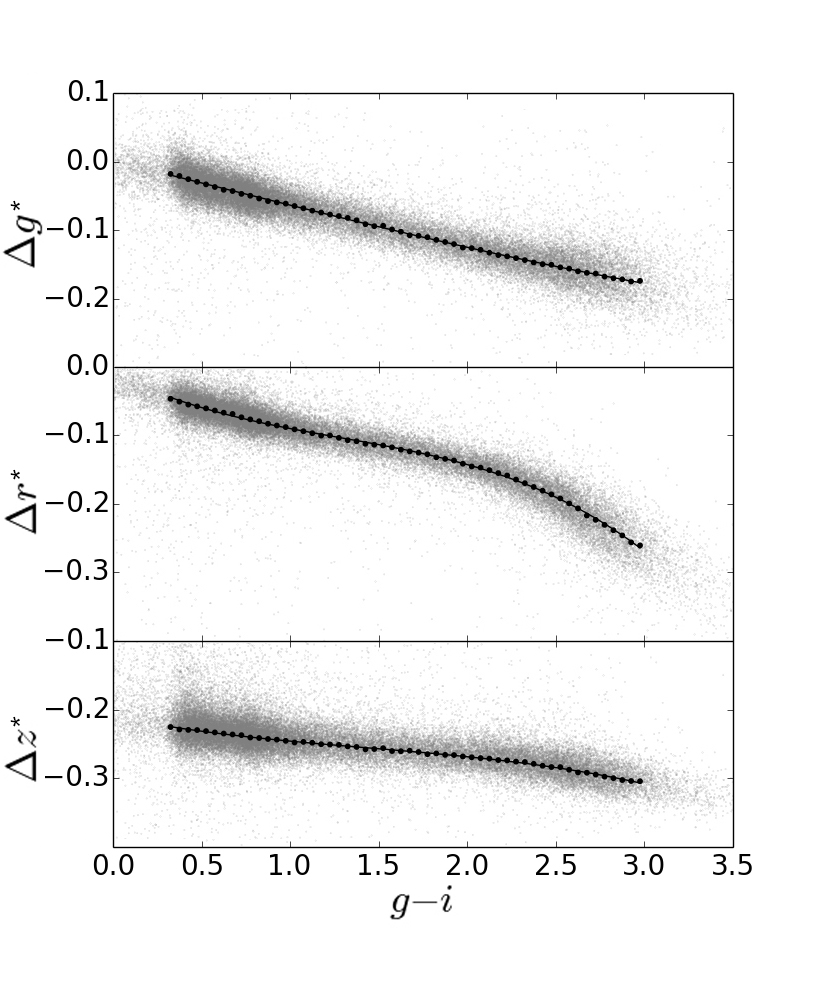}
\caption{The fitting results of the three bands describe the trend of mean values of the data. The gray dots represent all the data while the circles mark the mean values of $\Delta m^{\ast}$ in the bins. The solid lines show the fitting curves.}
\label{Figure_delta_gi}
\end{figure}

We note that SDSS uses $asinh$ magnitudes while DECaLS uses Pogson magnitudes. There will be a difference between these measures for very faint sources. Instead of directly using the difference of the magnitudes obtained from the catalog, the solution is to calculate the difference of fluxes as $\Delta m^{\ast} = -2.5\mathrm{log}(f_\mathrm{SDSS}/f_\mathrm{DECam}$) before the calibration, where $f$ represents flux and $\Delta m^{\ast}$ represents the magnitude difference. Furthermore, we also define the color $g-i$ in SDSS as $-2.5\mathrm{log}(f_g/f_i)$ in SDSS. We plot $\Delta m^{\ast}$ in the $g, r$, and $z$ bands vs. the color $g-i$ in SDSS in Figure \ref{Figure_delta_gi}. One way is fitting the binned data, instead of fitting all the data points. The $g-i$ color is divided into bins with a width of 0.05 and the bin centers are recorded, especially for all three bands of 0.3 $<g-i<$ 3.0, where the means of $\Delta m^{\ast}$ remain stable. Note that for the transformations between DECam and PS1\footnotemark[6], we apply a cubic fit and the accuracy of the cubic term is sufficient. In each bin we use $\sigma$-clipping to remove the outliers with the deviations from the mean values  greater than 10$\sigma$, where the standard deviation is determined by the Gaussian fit. Thus, the transformation formulas for converting from SDSS to DECam calculated with the standard stars are as follows ($g$ and $i$ represent the magnitudes in SDSS and the range of $g-i$ is 0.3 $<g-i<$ 3.0):
\begin{equation}
m_\mathrm{SD} = m_\mathrm{SDSS} + c_0 + c_1(g-i) + c_2(g-i)^2 + c_3(g-i)^3.
\label{Equation_transformation}
\end{equation}
The subscript ``SD'' refers to the DECam magnitude predicted from SDSS. The fitting curves are also shown in Figure \ref{Figure_delta_gi} and the coefficients are listed in Table \ref{Table_coefficient_transformation}. Compared to the present SDSS-PS1 transformation \citep[]{Morganson14} and PS1-DECam transformation, the results are close to the combination of the two transformations and the magnitude differences between them are $\lesssim$ 0.01 mag.

\begin{deluxetable}{c c c c c}
\tabletypesize{\scriptsize}
\tablecaption{Coefficients Used to Convert from SDSS Magnitudes to DECam Magnitudes in Equation (\ref{Equation_transformation})}
\tablewidth{0pt}
\tablehead{
\colhead{Filter} & \colhead{$c_0$} & \colhead{$c_1$} & \colhead{$c_2$} & \colhead{$c_3$}}
\startdata
$g$ & +0.00152 & $-$0.06464 & $-$0.00109 & +0.00091\\
$r$ & $-$0.00898 & $-$0.12964 & +0.06553 & $-$0.01707\\
$z$ & +0.01228 & $-$0.05673 & +0.02404 & $-$0.00544
\enddata
\label{Table_coefficient_transformation}
\end{deluxetable}

\subsection{The Photometric Noise Estimation}
\label{Section_2.3}

The remaining scatter in the magnitude differences after calibration is contributed by the photometric noise, providing that the standard stars are non-variable sources. However, the calibration stars are brighter than the quasars we are studying and thus have a larger typical signal-to-noise ratio (S/N). We must instead construct a reference sample of stars with measurements in SDSS and DECaLS that span the same magnitude range (and hence S/N) as the quasars.

\footnotetext[7]{http://legacysurvey.org/dr5/files/\#sweep-catalogs}
\footnotetext[8]{A DECaLS ``brick'' consists of a roughly $0.25 \times 0.25$ $\mathrm{deg^2}$ square region on the sky and is used to subdivide the survey area into smaller units.}
\footnotetext[9]{http://skyserver.sdss.org/dr12/en/tools/chart/listinfo.aspx}

We randomly choose 6 sweep files\footnotemark[7] where our quasars are contained, resulting in far more than 100 bricks\footnotemark[8]. Reference stars are selected by the following criteria: TYPE = ``PSF'' (morphological model) and NOBS\_[G, R, Z] = 1 (number of images that contribute to the central pixel in the $g$, $r$, and $z$ bands). In this way we create a reference star sample consisting of 48,575 stars. The SDSS information (magnitude, flux, and IVAR) of the reference stars can be directly searched out from the website\footnotemark[9]. Because these stars share the same bricks with quasars and only have single DECaLS observations, the photometric depths are similar with those of quasars. The reference stars can be considered as non-variable sources because the fraction of variable stars is very small and not likely to affect the statistics \citep{Sesar07}. This procedure assures that the S/N distribution of the reference stars is well matched with that of the quasars.

\begin{figure*}[]
\centering\includegraphics[width=1.0\linewidth]{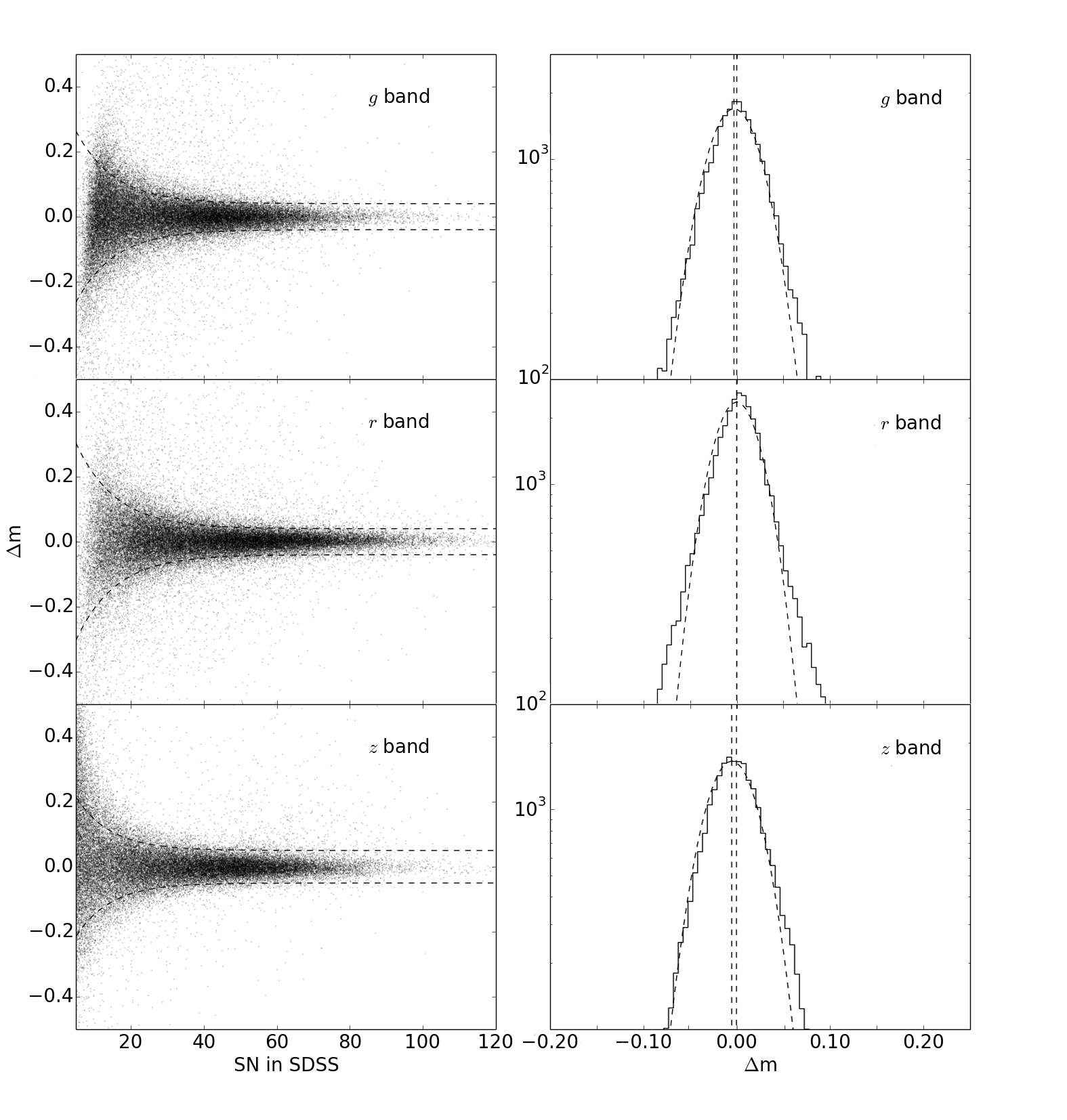}
\caption{($Left$) The magnitude differences after calibrations vs. S/N in SDSS in the $g (upper)$, $r (middle)$, and $z (lower)$ bands. The curves indicate the fitting results to the 68.3\% confidence half-width envelopes, which are symmetric about the y-axis. ($Right$) The distributions of magnitude differences after calibrations for the reference stars whose S/N is more than 25 in the $left$ panel in the $g (upper)$, $r (middle)$, and $z (lower)$ bands. The dashed lines indicate the Gaussian fits.}
\label{Figure_dm_SN_distribution}
\end{figure*}

As the SDSS-DECam magnitudes can be calculated using the transformation formulas in Equation (\ref{Equation_transformation}), the photometric noise is defined as the difference between the DECam magnitude and the SDSS-DECam magnitude for the reference stars 

\begin{equation}
\Delta m = m_\mathrm{DECam} - m_\mathrm{SD}.
\label{Equation_dm_definition}
\end{equation}
$\Delta m$ for the reference stars is the photometric noise, denoted as $\sigma_\mathrm{S/N}$ in our later analysis. Beware that $\sigma_\mathrm{S/N}$ here represents the magnitude difference for the non-variable reference stars to determine the photometric noise, rather than the photometric uncertainty, $\sigma_\mathrm{mag}$, which is directly measured in any of the two surveys.

We use S/N in SDSS to estimate the photometric noises because SDSS photometry is shallower than DECaLS. The mean S/N in the SDSS $g$, $r$, and $z$ bands is (38.9, 53.1, 42.1) while for DECam it is (298.9, 370.1, 624.4). For the reference stars we apply 

\begin{equation}
\sigma_\mathrm{S/N} = a_0 + a_1 \mathrm{exp} [a_2\times(\mathrm{S/N})/100]
\label{Equation_sigma_SN_definition}
\end{equation}
as our mathematical form of fitting the 68.3\% confidence half-width envelopes \citep{Vanden Berk04}. We notice that the fitting results are similar to Figure 2 in \cite{Vanden Berk04} and converge to 1$\sigma$ rapidly. Thus, we take the 1$\sigma$ values when S/N is larger than 25 (0.04, 0.04, and 0.05 mag) as the constant terms in the estimation for the $g$, $r$, and $z$ bands, respectively. The fitting results are shown in the left panel in Figure \ref{Figure_dm_SN_distribution} as well as in Table \ref{Table_coefficient_sigma_SN}.

\begin{deluxetable}{c c c c}
\tabletypesize{\scriptsize}
\tablecaption{Coefficients Used to fit the envelopes in Equation (\ref{Equation_sigma_SN_definition})}
\tablewidth{0pt}
\tablehead{
\colhead{Filter} & \colhead{$a_0$} & \colhead{$a_1$} & \colhead{$a_2$}}
\startdata
$g$ & $\pm$0.04 & $\pm$0.36 & $-$9.48 \\
$r$ & $\pm$0.04 & $\pm$0.42 & $-$9.27 \\
$z$ & $\pm$0.05 & $\pm$0.28 & $-$10.6
\enddata
\label{Table_coefficient_sigma_SN}
\end{deluxetable}

Another source of uncertainty is that the broad emission lines of quasars induce a different response to a given photometric system than the smooth continua of main-sequence stars. This may make the estimation imprecise. Thus, we additionally check the magnitude discrepancies between SDSS and DECaLS by simulating the quasar template \citep{Vanden Berk01} at different redshifts. The magnitude differences are calculated by convolving the template with the filter curves when redshift increases from 0.0 to 5.0 in steps of 0.01. For the $g$ and $r$ bands, over 67.3\% of the simulated differences are constrained to 0.01 mag. For the $z$ band, over 67.3\% of the differences is between 0.01 and 0.02 mag since these two $z$ bands are not very similar. We add additional 0.01 mag to our $z$-band photometric noise estimation.

\section{The Structure Function}
\label{Section_3}
We utilize the structure function (SF) to describe the ensemble quasar variability between SDSS and DECaLS. As given in Equation (12) of \cite{Kozlowski16r}, the expression is
\begin{equation}
V=\sqrt{\langle\Delta m^2-\sigma_\mathrm{S/N}^2\rangle},
\label{Equation_v_definition}
\end{equation}
where $\Delta m$ is the magnitude difference defined in Equation (\ref{Equation_dm_definition}) and $\sigma_\mathrm{S/N}$ is the photometric noise defined in Equation (\ref{Equation_sigma_SN_definition}). 1,573 quasars with magnitude differences greater than 1.0 are removed. While some of the large magnitude differences may be due to spurious measurements, e.g. on CCD edges, most are due to high-amplitude variability \citep[e.g.][]{Rumbaugh18}, which is beyond the scope of this study. \cite{Rumbaugh18} draw a conclusion that the large magnitude variances may be caused by the disk instabilities of low accretion rates, indicating a potentially different mechanism from the other quasars in our sample.

Equation (\ref{Equation_v_definition}) is the definition of the variability with the magnitude difference and the photometric noise, known as the SF. Here, $V$ describes the ensemble behavior of a set of quasars within each bin. An empirical model is derived in our study for the variance of the magnitude differences themselves, not the individual magnitudes. In this circumstance, our SF definition is not duplicated from Equation (12) in \cite{Kozlowski16r}.

As this is an ensemble study, we group magnitude differences for many quasars into bins of time lag, and then measure the SF within each bin. This provides an average measurement for the set of included quasars. We can then divide our full sample by quasar properties (redshift, bolometric luminosity, rest-frame wavelength, and black hole mass) to examine the dependence of the average variability on these properties. Because of two-epoch data, analyses on the quasar light curves are not applied in our work.

As in previous work, we adopt a power-law parameterization for the SF of quasar variability,
\begin{equation}
V(t | A, \gamma)=A(\frac{t}{1 yr})^\gamma.
\label{Equation_fitting_formula}
\end{equation}
In Section \ref{Section_4.1}, $A$ and $\gamma$ are constants and can be directly compared with previous work. In Section \ref{Section_4.3} and \ref{Section_4.4}, we discuss their dependency on quasar properties, that is, $A=A(z, L, \lambda, M)$ and $\gamma=\gamma(z, L, \lambda, M)$.

The SF is related to the auto-correlation function, $ACF(\Delta t)$ through the equation
\begin{equation}
SF(\Delta t)=SF_{\infty} \sqrt{1-ACF(\Delta t)}.
\label{Equation_SF_definition}
\end{equation}
This function may take the form of a power exponential (PE) as in \cite{Kozlowski17}:
\begin{equation}
ACF(\Delta t)=\mathrm{exp}[-(\frac{|\Delta t|}{\tau})^{\beta}].
\label{Equation_ACF_definition}
\end{equation}
Equation (\ref{Equation_SF_definition}) can be expanded into a Taylor series as $SF=SF_{\infty} (|\Delta t| {\tau}^{-1})^{\frac{\beta}{2}}$ when $|\Delta t| \ll \tau$, which is referred as the damped random walk (DRW) model when $\beta=1$. \cite{Mac10} found that $\tau$ is typically between 0.1 and 3 yr and $SF_{\infty}$ is typically between 0.1 and 0.5 mag. There are detailed discussions in \cite{Mac10} and \cite{Morganson14} about the relationship between DRW models and the ensemble SF. However, the conclusion is that the exponential model inherent to DRW cannot be easily applied to ensemble SFs. In reality, each quasar has its own set of SFs or DRW parameters. The ensemble SF treats all quasars as a single light curve and thus averages over the individual parameters. Because $SF_{\infty}$ is related to $\tau$ in the DRW model, one cannot robustly extract $SF_{\infty}$ without the knowledge of $\tau$ for each quasar. By averaging over a large number of objects, the results will tend toward a power-law relationship. In fact, a power-law fit is the short-term part of the general SF because we have little idea of how to determine the characteristic timescale $\tau$, that is, the turning point of the SF from a power-law form to a flat form. So, if we use a more complex damped random walk equation to fit the entire dataset, we will find the results almost identical to that of a power-law fit. Thus, we apply the power-law fit to quantify the analysis, considering the negligible difference between the two fits.

\section{QUASAR VARIABILITY AS A FUNCTION OF TIME LAG, REDSHIFT, BOLOMETRIC LUMINOSITY, REST-FRAME WAVELENGTH, AND BLACK HOLE MASS}
\label{Section_4}

\subsection{Rest-frame Time Lag}
\label{Section_4.1}

\begin{figure*}[]
\centering\includegraphics[width=1.0\linewidth]{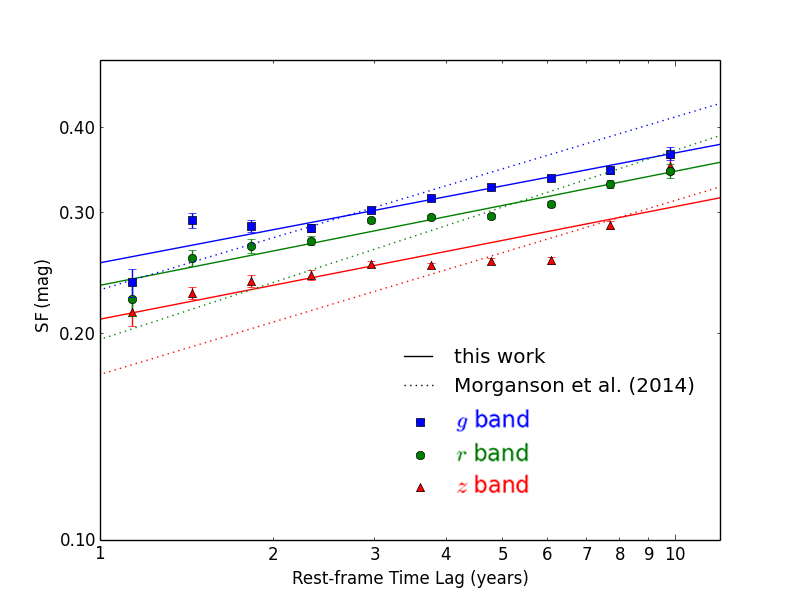}
\caption{The SFs for each of the three bands with single power-law fits to the bins are marked. The square dots indicate the variances in each bin and the solid lines indicate the single power-law fit results. In comparison we also plot the results of \cite{Morganson14} as the dotted lines. The blue, green, and red colors indicate the $g$, $r$, and $z$ bands, respectively.}
\label{Figure_SF_time_lag}
\end{figure*}

In this subsection we focus on the rest-frame time lag, $t=t_\mathrm{obs}/(1+z)$, in the three bands. This can be directly compared with the results in previous works. The time lag bins are divided into equal intervals on the logarithmic axis from 1 year to 11 years. The mean value of the two boundaries is taken. The rest-frame SFs for each of the three bands are presented in Figure \ref{Figure_SF_time_lag}. Rest-frame parameters including uncertainties in each of the bands are shown in Table \ref{Table_A_gamma}, along with the observer-frame SF to allow comparisons with those in \cite{Morganson14}. However, only in regard to the rest frame can we study the inherent characteristics of quasar variability. 

\begin{deluxetable*}{c c c c c c}
\tabletypesize{\scriptsize}
\tablecaption{Fitting Parameters of the Rest-frame and Observer-frame SFs in the Three Bands.}
\tablewidth{0pt}
\tablehead{
\colhead{Band} & \colhead{$N_{\rm quasars}$} & \colhead{Rest-frame $A$} & \colhead{Rest-frame $\gamma$} & 
\colhead{Observer-frame $A$} & \colhead{Observer-frame $\gamma$}}
\startdata
$g$ & 51,339 & 0.253$\pm$0.012 & 0.160$\pm$0.020 & 0.251$\pm$0.011 & 0.096$\pm$0.011\\
$r$ & 52,211 & 0.235$\pm$0.010 & 0.166$\pm$0.016 & 0.228$\pm$0.017 & 0.110$\pm$0.016\\
$z$ & 80,358 & 0.210$\pm$0.019 & 0.164$\pm$0.031 & 0.192$\pm$0.024 & 0.119$\pm$0.023
\enddata
\label{Table_A_gamma}
\end{deluxetable*}

In this study as well as the previous ones \citep[e.g.][]{Schmidt10}, the SF increases as a function of time lag. In addition, the variability amplitude decreases accordingly from the $g$ band to the $z$ band. This matches previous observations that the amplitude of variability decreases toward longer wavelengths \citep[e.g.][]{Vanden Berk01}. The shallower $z$-band data are compensated for by the larger number of observations in that band. We see sufficient quasars to mitigate the unreliability.

In the next three subsections we mainly concentrate on separating quasar properties and analyze the relationships of the variability with them. Notably, the first time lag bin tends to include more high-redshift quasars because of the definition of rest-frame time lag. The $g$-dropout effect may cause a significant redder $g-i$ color and correspondingly a lager magnitude offsets than low-redshift ones. We additionally check that in the first bin the fraction of the quasars at $z>3.7$ is only 10 percent, and the SF values will change no greater than 0.01 mag when removing these $g$-dropout quasars.

\subsection{Multi-dimensional Fit and Bootstrap Method}
\label{Section_4.2}

We now examine the relationship between variability and other properties of quasars, namely redshift, bolometric luminosity, rest-frame wavelength, and black hole mass. Rest-frame wavelength is defined as $\lambda=\lambda_\mathrm{obs}/(1+z)$, where $\lambda_\mathrm{obs}$ is given by the central wavelength of the SDSS $g, r$, and $z$ filters (4686\AA, 6166\AA, and 8932\AA). To study the variability as a function of these properties, each of them is limited to a small range. As a result, we divide redshift, bolometric luminosity, rest-frame wavelength, and black hole mass each into 5 bins. Considering that our quasar population is not evenly distributed in redshift, DR12Q is composed of more quasars at $2.0 < z < 3.0$, we divide the properties so that the number of quasars in each bin is close to one-fifth of the whole population. The redshift bins are bounded at z = 1.07, 1.94, 2.33, and 2.65. The bolometric luminosity bins are bounded at $L$ = 45.71, 46.03, 46.28, and 46.56 $\mathrm{erg \cdot s^{-1}}$. The rest-frame wavelength bins are bounded at $\lambda$ = 1600, 2120, 2630, and 3360 \AA. The black hole mass bins are bounded at $M_\mathrm{BH}$ = 8.39, 8.71, 8.96, and 9.23 $\mathrm{M_{\odot}}$. The rest-frame time lag bin boundaries are mentioned above. Thus, we get 6,250 cells (we use bins for 1-dimensional data and cells for multi-dimensional data): 625 cells in a $z$-$L$-$\lambda$-$M_\mathrm{BH}$ 4-dimensional space along with additionally 10 bins of the rest-frame time lag. The large quasar sample makes it possible to separate the data into so many cells. We reject the cells containing fewer than 10 quasars. Since the quasar population is not evenly distributed in the the $z$-$L$-$\lambda$-$M_\mathrm{BH}$ 4-dimensional space, there will be some cells containing few quasars. Therefore, the number of cells decreases finally to $\sim$1,600. We use the mean value of redshift, bolometric luminosity, rest-frame wavelength, and black hole mass of the quasar population to represent a whole cell.

We start by examining how the variability parameters depend on the parameters above. Thus, from Equation (\ref{Equation_fitting_formula}) we adopt Equation (19) in \cite{Morganson14} for the cells described above:

\begin{equation}
\begin{array}{ccc}
A = A_0 (1+z)^{B_z} (\frac{L}{L_{46}})^{B_L} (\frac{\lambda}{\lambda_4})^{B_{\lambda}} (\frac{M}{M_9})^{B_M},\\
\\
\mathrm{log}A = \mathrm{log}A_0 + B_z\mathrm{log}(1+z) + B_L\mathrm{log}(L/L_{46}) \\
+ B_{\lambda}\mathrm{log}(\lambda/\lambda_4) + B_M\mathrm{log}(M/M_9),\\
\\
\gamma = \gamma_0 + \beta_z z + \beta_L\mathrm{log}(L/L_{46}) \\
+ \beta_{\lambda}\mathrm{log}(\lambda/\lambda_4) + \beta_M\mathrm{log}(M/M_9), \\
\\
L_{46} = 10^{46} \mathrm{erg \cdot s^{-1}}, \lambda_4 = 10^4 \mathrm{\AA}, M_9 = 10^9 \mathrm{M_{\odot}}.
\end{array}
\label{Equation_multidimensional_fit}
\end{equation}

A multi-dimensional fit is applied, considering the reciprocal of the error bar as a weight. Compared to the methods in \cite{Morganson14}, we add both rest-frame wavelength and black hole mass simultaneously into the equations.

\begin{figure*}[]
\centering\includegraphics[width=1.0\linewidth]{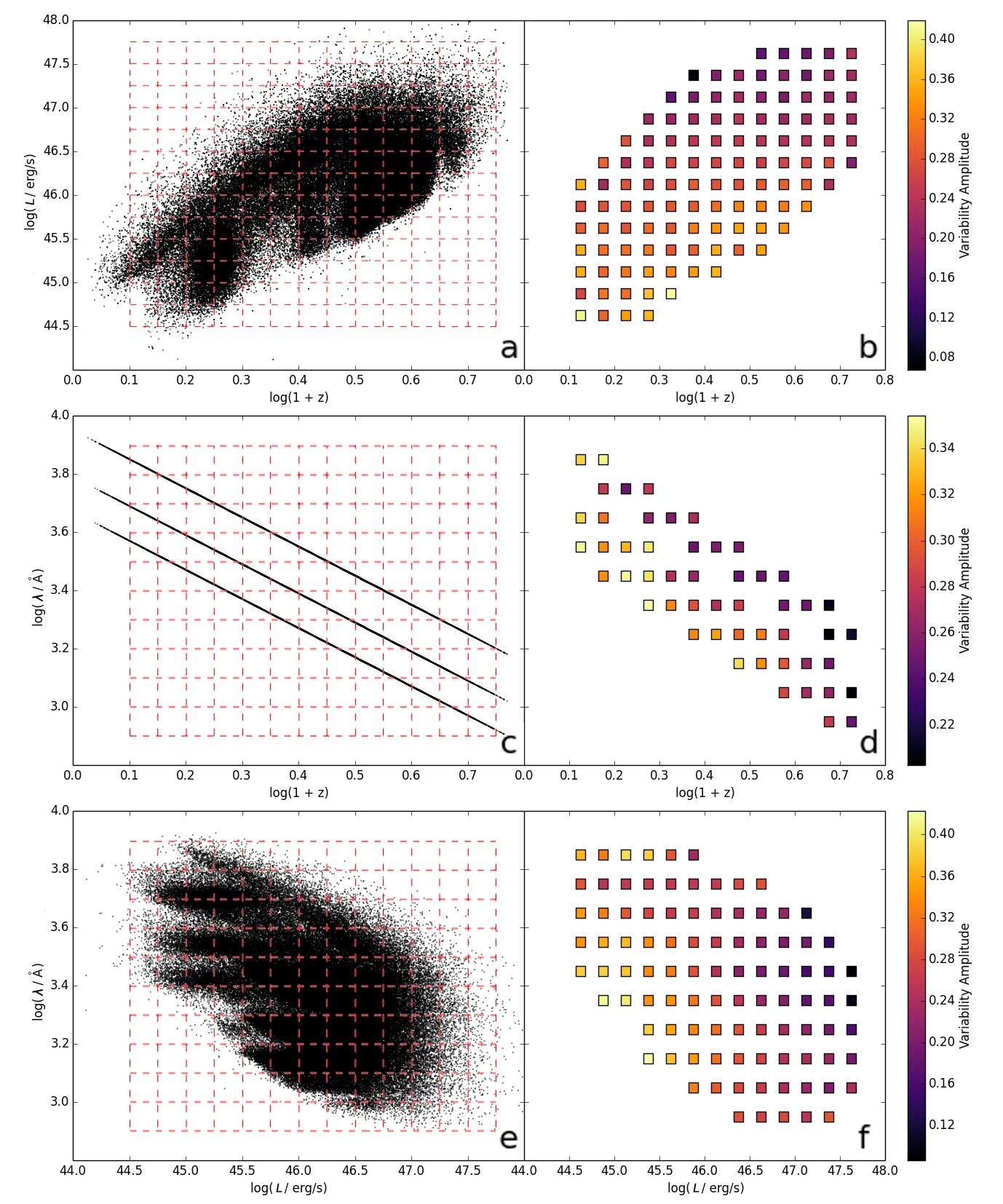}
\caption{From top to bottom the $left$-$right$ pairs represent $z$-$L$ space, $z$-$\lambda$ space and $L$-$\lambda$ space, respectively. The black dots represent the quasars. The red grids are not the bin boundaries in Section \ref{Section_4.2}. We can divide the sample into more cells in the color bar figures because they contain only two properties. It is clear that a couple of cells contain few quasars. The reason is similar to that described in Section \ref{Section_4.2} why the number of cells decreases from 6,250 to $\sim$1,600.}
\label{Figure_zLl}
\end{figure*}

In addition, we apply a bootstrap method to quantify the significance of the fitted values. The initial quasar data are resampled for 500 times and repeatedly fitted. This procedure makes the number of total cells unfixed, fluctuating between 1,572 and 1,633. One standard deviation is adopted to present the tolerance of one physical quantity. Thus, we obtain:

\begin{equation}
\begin{array}{cccccc}
A_0 = 0.076\pm0.010 , \gamma_0 = 0.487\pm0.060 , \\
\\
B_z = 0.286\pm0.123 , \beta_z = -0.050\pm0.028 , \\
\\
B_L = -0.251\pm0.021 , \beta_L = 0.089\pm0.034 , \\
\\
B_{\lambda}=-0.451\pm0.055 , \beta_{\lambda} = 0.172\pm0.089 , \\
\\
B_M = -0.019\pm0.018 , \beta_M = 0.168\pm0.029.
\end{array}
\label{Equation_fitting_result}
\end{equation}
The variations in the parameter fits are not as large as the parameters, indicating that the results are statistically significant.

\subsection{The Variability Amplitude Dependency on the Quasar Properties}
\label{Section_4.3}

We next focus on the variability amplitude utilizing color bar figures which contain only two properties. Figure \ref{Figure_zLl} shows the 2-dimensional color bar relations in a more direct way. This method also avails us to disentangle the influence of several properties. Similarly, we reject the cells containing 25 quasars or fewer. Within each cell we only illustrate the ensemble variability defined in Equation (\ref{Equation_v_definition}).

In Panels a, b, e, and f in Figure \ref{Figure_zLl} the well-known negative relation of the variability amplitude with luminosity is clearly seen. This anti-correlation is statistically significant in view of Equation (\ref{Equation_fitting_result}).

In Panels c and d in Figure \ref{Figure_zLl}, because $\lambda$ is calculated by dividing the central wavelength by the constant, three straight lines appear in log-log space. The negative relation of the variability amplitude with rest-frame wavelength is also indicated in Table \ref{Table_A_gamma}, as the $g$-band variance is greater than the others. According to the standard thin-disk model of quasar accretion disks \citep{Shakura76}, temperature decreases with increasing radius. \cite{Mac10} thus interpreted the wavelength dependence of variability as a lower variability amplitude at larger radii.

$B_z$ is found to be positive, which suggests an increase in variability with redshift even at fixed wavelength. However, the significance of the $B_z$ parameters is only $\sim2\sigma$, indicating low confidence in this result. Generally, the bluer part tends to be more variable than the redder part, so the increasing variability amplitude appears to depend on an increasing redshift \citep{Morganson14}.

As $B_M$ is only a $\sim1\sigma$ result, the variability amplitude dependency on black hole mass is still uncertain, and we do not include black hole mass in Figure \ref{Figure_zLl}.

\subsection{$\gamma$ Dependency on the Quasar Properties}
\label{Section_4.4}

\begin{figure}[]
\centering
\includegraphics[width=0.9\linewidth]{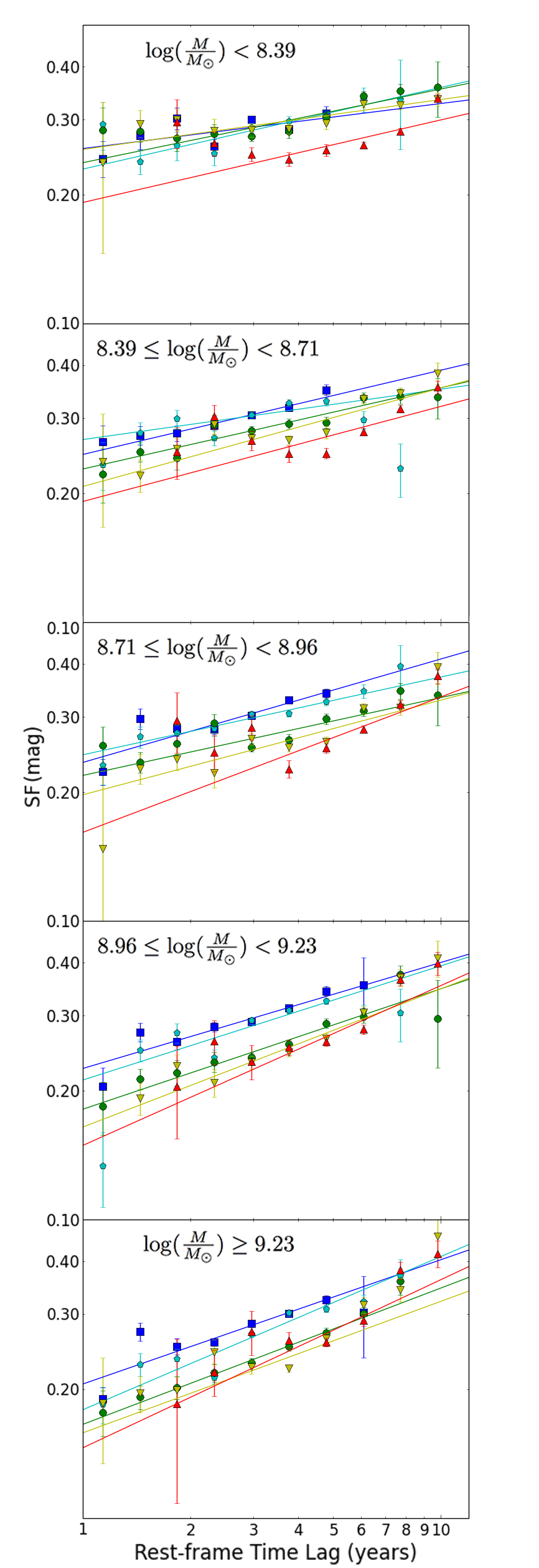}
\caption{The SFs at fixed black hole masses and for different rest-frame wavelengths. The blue, cyan, green, yellow, and red colors represent the rest-frame wavelengths at $<$1600\AA, 1600\AA-2120\AA, 2120\AA-2630\AA, 2630\AA-3360\AA, and $\ge$3360\AA, respectively.}
\label{Figure_gamma_dependency}
\end{figure}

The $\gamma$ value illustrates how the SF increases with time lag. Noticing the fact that it is only slightly more than a $2\sigma$ significance, the rest-frame wavelength dependence of $\gamma$ need to be justified more carefully. Figure \ref{Figure_gamma_dependency} leads to visualize how strong the evidence is for $\gamma$ varying in different wavelength bins. We separately plot the rest-frame SFs for the 5 rest-frame wavelength bins and the 5 black hole mass bins. The rest-frame wavelength bins and the black hole mass bins are mentioned in Section \ref{Section_4.2}. It suggests that the slope is larger with an increasing black hole mass. However, for the rest-frame wavelength, the slopes look noisy, which is consistent with the $\beta_{\lambda}$ term only having $\sim2\sigma$ significance. It appears that most of the wavelength dependence may be coming from the fact that the red points ($\lambda>3360$\AA) at $t>7$ years are consistently high, which mimics the $z$-band behavior of the SF in Figure \ref{Figure_SF_time_lag}. In other words, we presume that there is some systematic issue driving the $z$-band points up, which is then driving a mild apparent steepening of $\gamma$ with respect to the rest-frame wavelength. As for luminosity dependence, the relation can also be explained by Figure \ref{Figure_SF_time_lag}: our SF fitting slopes are flatter in all the three bands than those in \cite{Morganson14}. It is partly because our sample includes both DR12Q and DR7Q while only the latter is considered in \cite{Morganson14}. Assuming that the relation of $\gamma$ with redshift is negligible as reported, the main difference is luminosity. DR12Q has more faint quasars than DR7Q (See Figure \ref{Figure_L_z_distribution}). According to the positive $\beta_L$ value in Equation (\ref{Equation_fitting_result}), the lower luminosity is, the lower the $\gamma$ value is.

Notably, we find that $\gamma$ is partly dependent on the quasar properties, not as \cite{Morganson14} reported, who first dealt with redshift and luminosity, then fit rest-frame wavelength, assuming $\gamma$ to be a constant. We notice $\gamma$ is significantly associated with bolometric luminosity, rest-frame wavelength and black hole mass and is weakly correlated with redshift. 

\begin{figure}[]
\centering
\includegraphics[width=1.0\linewidth]{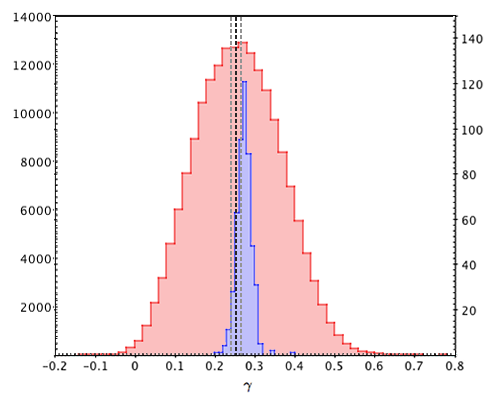}
\caption{The red semi-step distribution: the recovered $\gamma$ values from each quasar from our fitting procedure with the properties ($\gamma=0.259\pm0.106$, 183,908 values of 119,305 quasars in total, 51,339 in the $g$ band, 52,211 in the $r$ band, and 80,358 in the $z$ band, see the y-axis on the left); the black and gray dashed lines: the outcome of fitting the original data considering $\gamma$ as a constant ($\gamma=0.254\pm0.011$); the blue semi-step distribution: the outcomes of fitting the resampled data by the bootstrap method considering $\gamma$ as a constant ($\gamma=0.274\pm0.019$, 500 times in total, see the y-axis on the right).}
\label{Figure_gamma_distribution}
\end{figure}

However, a typical $\gamma$ value cannot be easily derived because Equation (\ref{Equation_fitting_result}) only depicts how the $\gamma$ value may vary with the quasar properties. In order to determine a specific $\gamma$ value of our quasar sample, we try the following two ideas. First, we test how $\gamma$ is distributed as a function of these four properties. We recover the $\gamma$ value from each quasar from our fitting procedure and Figure \ref{Figure_gamma_distribution} clarifies that it is approximately and larger than 0.25. The average is 0.259 and the standard error is 0.106. Second, $\gamma$ can again be estimated by being assumed a constant, $\gamma = \gamma_0$. We aim to get a fixed value if $\gamma$ does not depend on the quasar properties, while $A$ is still a function of the quasar properties. Following the way of \cite{Morganson14}, we modify the basic fitting formula to $V=A(t/1yr)^{\gamma_0}$, where $A$ can be expressed the same as the first formula in Equation (\ref{Equation_multidimensional_fit}) while $\gamma_0$ no longer serves as a function of quasar properties. In this circumstance, we can directly fit the data again, and the outcome is $\gamma_0=0.254\pm0.011$. We can also first apply the bootstrap method, then fit the resampled data each (500 times in total). Such 500 outcomes lead to a mean value and a standard deviation, which are 0.274 and 0.019 (see Figure \ref{Figure_gamma_distribution}). This is consistent with the results reported above and larger than that in \cite{Morganson14}.

\section{THE VARIABILITY AMPLITUDE DEPENDENCY ON THE EDDINGTON RATIO}
\label{Section_5}

\subsection{An Extended Result of Black Hole Mass}
\label{Section_5.1}

We now focus specifically on black hole mass and the Eddington ratio. Since we directly use the black hole masses of DR12Q estimated by \cite{Kozlowski16}, we examine the effect of the quasar broad emission lines on the flux and magnitude measurements. In this work the empirical correction ratio $R$ is used to convert the broad-band magnitudes to the monochromatic luminosities. This correction is fitted with DR7Q with a typical dispersion of $\approx$ 0.1 dex. Eventually we decide to quote this set of data.

\cite{Kozlowski16} provided the estimated monochromatic luminosities and the black hole masses at 5100\AA (H$\beta$), 3000\AA($\mathrm{Mg_{II}}$), and 1350\AA ($\mathrm{C_{IV}}$), respectively. Because the black hole masses estimated by the emission lines H$\beta$ and $\mathrm{Mg_{II}}$ are more accurate than that of $\mathrm{C_{IV}}$ whose dynamics may be strongly influenced by outflows \citep{Willott03}, we take the H$\beta$ and $\mathrm{Mg_{II}}$ as higher priority to calculate the black hole masses.
 
The relationship of the variability amplitude with both black hole mass and the Eddington ratio are shown in Figure \ref{Figure_MR}. It has been debated for a long time whether black hole mass is a key attribute of the quasar variability amplitude. Our results show that black hole mass has an impact on the variability amplitude, but the exact link is still unclear. The relationship of the variability amplitude with black hole mass seems negative while an upward trend of black hole mass at fixed luminosity is also found. This is also demonstrated by the multi-dimensional fitting result with the variability amplitude and black hole mass. As we mentioned before, the index, $B_{M}$, is very low compared with its standard deviation, indicating that the overall variability amplitude is only weakly a function of black hole mass. As a result, we prudently report that the black hole mass relation is still uncertain and needs further investigations.

It is possible to propose that the greater black hole mass is, the less the quasar is influenced by a perturbation, no matter what causes the variability. However, another concept has been proposed to explain why the variability amplitude is correlated with black hole mass: the more massive black holes are starving and providing large flux variability because they do not have a steady inflow of gaseous fuel \citep{Wold07}.

\subsection{The Eddington Ratio as a Main Factor of Variability}
\label{Section_5.2}

The Eddington ratio is proposed to reflect the extent of growth compared with bolometric luminosity. By definition, the Eddington ratio is determined by $R_{\mathrm{Edd}} = L / L_{\mathrm{Edd}}$. We notice that $L_{\mathrm{Edd}} \propto M_{\mathrm{BH}}$, and $R_{\mathrm{Edd}} \propto L / M_{\mathrm{BH}}$ is given, where $R_{\mathrm{Edd}}$ represents the Eddington ratio, $L$ is bolometric luminosity and $L_{\mathrm{Edd}}$ is Eddington luminosity. In addition, given that $M_{\mathrm{BH}}$ is estimated by $M_{\mathrm{BH}} \propto L^{1/2} \mathrm{FWHM}^2$, the statement $R_{\mathrm{Edd}} \propto L^{1/2} / \mathrm{FWHM}^2$ is sound, where FWHM is short for full width at half maximum for a specific kind of emission lines (e.g. $\mathrm{H}\mathrm{\beta}$). Thus, if at the moment we do not see the cause of such changes, we are not confident enough to determine whether luminosity or the Eddington ratio serves as a more influential factor.

\begin{figure*}[]
\centering
\includegraphics[width=1.0\linewidth]{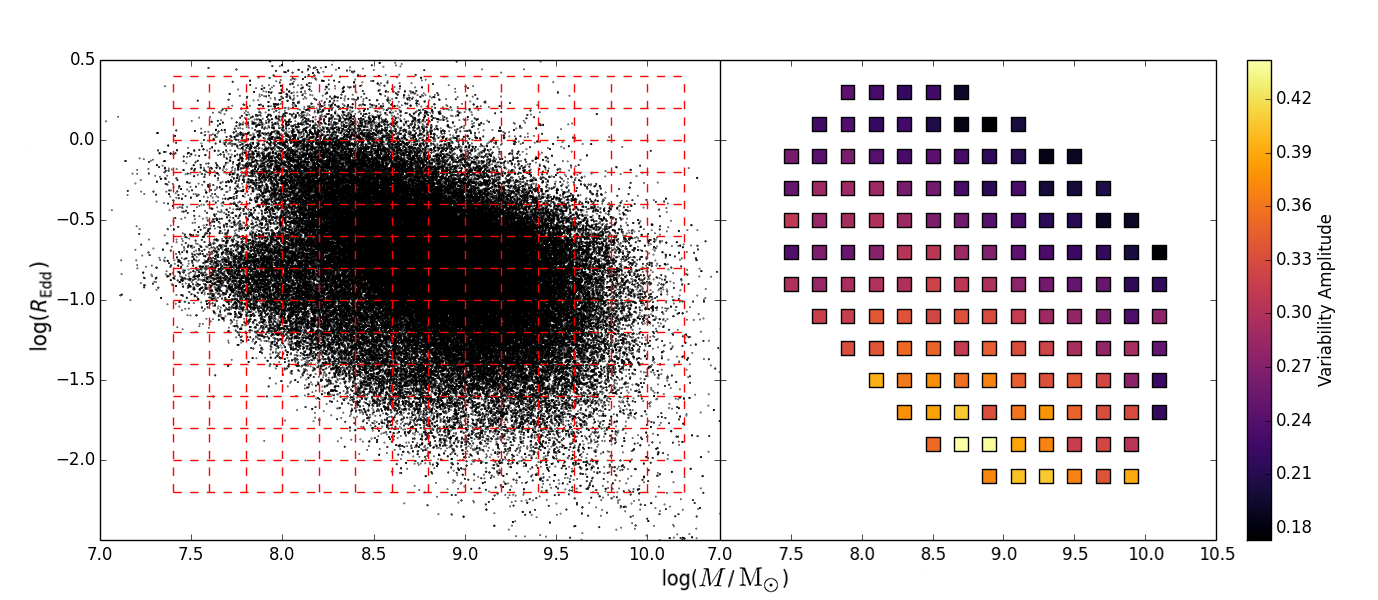}
\caption{The variability amplitude as a function of both black hole mass and Eddington ratio. The Eddington ratio contributes obviously to the variability amplitude while black hole mass influences is less clear.}
\label{Figure_MR}
\end{figure*}

For the amplitude dependence, we expect that the Eddington ratio plays an important role because it is clear in Equation (\ref{Equation_fitting_result}) that luminosity dependence is negative while black-hole-mass dependence is negligible. We compare the Eddington ratio and black-hole-mass dependence by rescaling the exponents in Equation \ref{Equation_fitting_result} without the black-hole-mass ones. The variability amplitude is significantly anti-correlated with the Eddington ratio, as shown in Figure \ref{Figure_MR}. However, we still have to mention that the black hole mass estimates and the Eddington ratio estimates have a $\sim$0.3 dex scatter. Since the error of black hole mass or the Eddington ratio is not given in the DR12 quasar catalog provided by \cite{Kozlowski16}, we do not take them into consideration.

Again, the index in Equation (\ref{Equation_fitting_result}) also confirms black hole mass as a weaker influencing factor than the Eddington ratio. Compared with black hole mass, the variability amplitude can be described as a function that strongly depends on both luminosity and the Eddington ratio. This is the same trend as that reported in the previous work of \cite{Mac10}, who argued that the Eddington ratio may be the primary driver of the variability amplitude.

\section{DISCUSSION}
\label{Section_6}

The DECaLS-SDSS dataset makes it possible to research on the quasars variability with a wide range of quasar properties. Compared with \cite{Morganson14}, \cite{Zuo12}, \cite{Mac10} and several other studies, our sample contains more quasars, especially the fainter ones and reaches a higher redshift.

For the dependency of the variability amplitude on the rest-frame time lag, the relationships are quite clear in the references above \citep[][]{Morganson14, Vanden Berk04}. A power-law index of approximately $\gamma = \beta/2 \gtrsim 0.25$ is clearly discovered in our study. Our index is close to but slightly higher than 0.2457 \citep{Morganson14} and 0.246 \citep{Vanden Berk04}. However, \cite{Kozlowski16r} noted that the definition of the SF in \cite{Vanden Berk04} subtracts an incomplete noise term and leads to flatter SFs. Our study provides a correspondingly ``steeper'' results with the combination of both the multi-dimensional fit and the bootstrap method. \cite{Bauer09} provided a table (Table 4) listing their results as well as the previous ones. However, their quasar sample is smaller compared with ours.

As for the dependency of the variability amplitude on bolometric luminosity, an anti-correlation is proposed in our study. A quantity of work presented the same negative relation \citep{Vanden Berk04, Mac10, Zuo12}. Simply focusing on how the variability amplitude depends on bolometric luminosity, our logarithm index value of $-$0.251 is similar to $-$0.200 \citep[][]{Morganson14} and $-$0.205 \citep[][]{Bauer09}.

Combining Figure \ref{Figure_SF_time_lag} and the multi-dimensional fit, we confirm the well-known bluer-when-brighter trend. The variability amplitude decreases at longer wavelengths with a power-law index of $-$0.451. The same conclusion was also reached by \cite[][$-$0.44]{Morganson14}, \cite[][$-$0.479]{Mac10}, and \cite{Zuo12}. On the other hand, the redshift dependence is less clear since we find a positive trend but it is not statistically significant ($2\sigma$) in view of the later applied bootstrap method. \cite{Vanden Berk04} presented different formula expressions to fit redshift and wavelength, which are respectively described as a straight upward linear function and a downward exponential function. \cite{Zuo12} showed no relation with redshift and \cite{Mac10} reported a negligible trend with redshift. \cite{Morganson14} provided a positive relation with a power-law index of 0.153 along with an interpretation of a joint effect of redshift-wavelength combination. Note that our quasar sample reaches a maximum redshift of 4.89. This, along with the considerable quasar population thanks to DECaLS survey, enables us to check for more complete relations.

We take black hole mass into consideration while \cite{Morganson14} did not. We provide a $\sim 1\sigma$ relation of the variability amplitude with black hole mass, which means that no specific relation is discovered. This supports the statements in \cite{Zuo12}, but is not consistent with the results provided by \cite{Mac10} and \cite{Bauer09}.

Our study provides a series of relations of the variability increasing rate, $\gamma$, with the quasar properties mentioned above. No clear relation of $\gamma$ with redshift is discovered in our study. We clarify the positive trends with black hole mass by showing Figure \ref{Figure_gamma_dependency} for different rest wavelengths in each black hole mass bin. However, the seemingly positive rest-frame wavelength dependence cannot be easily determined based on the bootstrap result along with the noises in Figure \ref{Figure_gamma_dependency}. The correlation of $\gamma$ with luminosity is confirmed, which explains the phenomenon that our SF fitting slopes are flatter than those in \cite{Morganson14} by considering the $\gamma$ dependency on bolometric luminosity. This result is consistent with the empirical relation proposed by \cite{Kozlowski16r}, who reported $\beta \varpropto L^{0.1}$ derived from $\sim9,000$ light curves in SDSS Stripe 82.

\begin{deluxetable*}{c c c c c}
\tabletypesize{\scriptsize}
\tablecaption{The Relationships of Variability Parameters with Quasar Properties}
\tablewidth{0pt}
\tablehead{
\colhead{Variability Parameters} & \colhead{Redshift} & \colhead{Bolometric luminosity} & \colhead{Rest-frame wavelength} & \colhead{Black hole mass}}
\startdata
$A$ & positive ($2\sigma$) & negative & negative & uncertain\\
$\gamma$ & negative ($2\sigma$) & positive ($2\sigma$) & positive ($2\sigma$) & positive
\enddata
\tablecomments{All the relationships discussed in Section \ref{Section_4.3} and \ref{Section_4.4} except the Eddington ratio. The bootstrap method reliability is shown in the parenthesis if the relation of a variability parameter with one property is not greater than $3\sigma$.}
\label{Table_results}
\end{deluxetable*}

Although the microlensing hypothesis provides $\gamma$ = 0.23 to 0.31, which is consistent with our result, it still remains a minor cause for the sake of the following two reasons. First, microlensing activity should be significantly rare at low redshift. The variability amplitude should correspondingly decrease when redshift decreases. The redshift dependence in our study does not reveal a robust correlation with redshift. Second, microlensing model are not related with the intrinsic accretion activity, and cannot explain the obvious correlation with the Eddington ratio. \cite{Kawaguchi98} predicted that power-law form slope should range from 0.41 to 0.49 in view of disk instability. However, it is a simplified model because of the assumptions of the intrinsic accretion activity. More quantitatively theoretical predictions on the other quasar properties need to be compared with our work.

\section{CONCLUSIONS}
\label{Section_7}

The SDSS-DECaLS dataset gives us an opportunity to apply several methods to examine the quasar ensemble variability with reasonable precision. Within over $10^5$ quasars and the deep detection of DECaLS, we test the quasar ensemble variability dependency on the quasar properties. Considering the rest-frame time lag, redshift, bolometric luminosity, rest-frame wavelength, and black hole mass, we fit the entire dataset to determine the dependency of both the variability amplitude and the variability increasing rate on quasar properties. Our main conclusions are as follows:

1. We use a recent well-established definition of SF \citep{Kozlowski16r} and a single power-law $V=A(t/1yr)^{\gamma}$ to fit the binned data. Our analysis provides $\gamma = \beta/2 \gtrsim 0.25$. The similar but flatter results were also reported in \cite{Vanden Berk04} and \cite{Morganson14}.

2. We confirm the anti-correlation of the variability amplitude with bolometric luminosity, which was clarified by a number of previous studies. The negative relation of the variability amplitude with rest-frame wavelength is clarified. The bluer parts tend to be more variable than the redder parts. The relation with redshift could be interpreted as a simultaneous effect as proposed by \cite{Morganson14}, but only an unstable correlation is confirmed in this work. Notably, we report the relations of the variability increasing rate, $\gamma$, with the quasar properties. Specifically, $\gamma$ depends more significantly on bolometric luminosity (correlated), rest-frame wavelength (correlated, further investigations are needed) and black hole mass (correlated) than on redshift (anti-correlated, further investigations are needed). We confirm the correlation of $\gamma$ with black hole mass, and explain the correlation with luminosity from the comparison with \cite{Morganson14}. The trends are also illustrated by the multi-dimensional fit and the bootstrap method analysis. However, our sample is too noisy to determine the relationships with redshift and the rest-frame wavelength. This part can also be found in Table \ref{Table_results} as well as some indications of bootstrap method reliability in the corresponding parentheses, which illustrate the comparisons between the results and the standard deviations.

3. We compare both the microlensing model and the disk-instability model for quasar variability. Although the microlensing model predicts a $\gamma$ range which is consistent with our result, the disk-instability model is more promising when discussing some intrinsic causes of our results. Thus, our study favors the latter.

Thanks to DECaLS and SDSS being two fairly successful surveys, we have established a large dataset of quasars. Redshift range makes our results universal. Compared to DECaLS, SDSS photometric accuracy is relatively low and correspondingly the total uncertainty is affected. We control the photometric uncertainty to the desired range by expanding the number of quasar samples. At the same time, it also reminds us to use more survey projects with either deep fields or larger quasar populations to obtain more reliable data.

In the future, DECaLS will release the multi-epoch data enabling better analysis because the multi-epoch data can better define the light curve. Maximum likelihood will probably be applied at that time. 

Finally, as a by-product, we are able to select a couple of changing-look quasars if we focus on extremely variable ones. Magnitude variance $>$ 2.5 is one of the criteria. We are looking for a kind of changing-look quasars whose continuum shows up in SDSS but disappears in DECaLS. This may open a door to understand the nature of broad line region of quasars.

\acknowledgments{We thank the supports by the NSFC grant No.11373008 and 11533001, the National Key Basic Research Program of China 2014CB845700, and from the Ministry of Science and Technology of China under grant 2016YFA0400703. We thank Fuyan Bian, Jinyi Shangguan, Wenwen Zuo, Jinyi Yang, Feige Wang, and Bingxu Yao for helping revising it. We also acknowledge to Christian Wolf, Szymon Koz\l{}owski, Matthew A. Malkan, Robert J. Brunner, and Donald G. York for helpful discussions.}

\footnotetext[10]{http://portal.nersc.gov/project/cosmo/data/legacysurvey/dr3 \\ /survey-ccds-decals.fits.gz}

\appendix{
MJD is not available in the files mentioned in Section \ref{Section_2.1}. We use DECaLS CCD files in DR3\footnotemark[10] containing previous DR1 and DR2 information. Here, we apply a special 2-d Cuboid cross-match to find which CCD detects the individual quasar. Once the quasar coordinates lie in the boundary limit (x error = 0.140 and y error = 0.075 from the center) of one CCD, it is considered to be detected by this CCD. We note that there are considerable quasars marked as non-DECaLS, meaning that they cannot be found in any CCDs in our CCD file. Keeping in mind that one quasar may be observed many times, we also notice that DR3 only contains one average flux in the file. Since we must consider those quasars that were observed in a single day, if one was observed again several days or months later, its flux could not give us accurate information. By checking the CCD files, we obtain the MJD information in DECaLS.
}


\begin{thebibliography}{}

\bibitem[Ai et al.(2010)]{Ai10} Ai, Y. L., Yuan, W., Zhou, H. Y., et al. \ 2010, \apj, 716, L31
\bibitem[Angione \& Smith(1972)]{Angione72} Angione, R. J., \& Smith, H. J. 1972, in IAU Symp. 44, External Galaxies and Quasi-Stellar Objects, ed. D. S. Evans, D. Wills, \& B. J. Wills (Dordrecht: Reidel), 171
\bibitem[Aretxaga et al.(1997)]{Aretxaga97} Aretxaga, I., Cid Fernandes, R., \& Terlevich, R. J. \ 1997, \mnras, 286, 271
\bibitem[Bauer et al.(2009)]{Bauer09} Bauer, A., Baltay, C., Coppi, P., et al. \ 2009, \apj, 696, 1241 1, 5.1
\bibitem[Caplar et al.(2016)]{Caplar16} Caplar, N., Lilly, S. J., Trakhtenbrot, B., \ 2017, \apj, 834, 111
\bibitem[Chang \& Refsdal(1979)]{Chang79} Chang, K., \& Refsdal, S. \ 1979, \nat, 282, 261
\bibitem[Cid Fernandes, Aretxaga \& Terlevich(1996)]{Cid96} Cid Fernandes, R., Jr., Aretxaga, I., \& Terlevich, R. \ 1996, \mnras, 282, 1191
\bibitem[Cid Fernandes et al.(1997)]{Cid97} Cid Fernandes, R., Terlevich, R., \& Aretxaga, I. \ 1997, \mnras, 289, 318
\bibitem[Collier \& Peterson(2001)]{Collier01} Collier, S., \& Peterson, B. M. \ 2001, \apj, 555, 775
\bibitem[Courvoisier et al.(1996)]{Courvoisier96} Courvoisier, T. J.-L., Paltani, S., \& Walter, R. \ 1996, \aap, 308, L17
\bibitem[Cristiani et al.(1996)]{Cristiani96} Cristiani, S., Trentini, S., La Franca, F., et al. \ 1996, \aap, 306, 395
\bibitem[Cutri et al.(1985)]{Cutri85} Cutri, R. M., Wisniewski, W. Z., Rieke, G. H., \& Lebofsky, M. J. \ 1985, \apj, 296, 423
\bibitem[Czerny et al.(2008)]{Czerny08} Czerny, B., Siemiginowska, A., Janiuk, A., \& Gupta, A. C. \ 2008, \mnras, 386, 1557
\bibitem[de Vries et al.(2003)]{de03} de Vries, W. H., Becker, R. H., \& White, R. L. \ 2003, \aj, 126, 1217
\bibitem[Dexter \& Agol(2010)]{Dexter10} Dexter, J., \& Agol, E. \ 2010, ascl.soft, 11015
\bibitem[di Clemente et al.(1996)]{di96} di Clemente, A., Giallongo, E., Natali, G., Tr\`evese, D., \& Vagnetti, F. \ 1996, \apj, 463, 466
\bibitem[Doi et al.(2010)]{Doi10} Doi, M., Tanaka, M., Fukugita, M., et al. \ 2010, \aj, 139, 1628
\bibitem[Garcia et al.(1999)]{Garcia99} Garcia, A., Sodr\'e, L., Jr., Jablonski, F. J., \& Terlevich, R. J. \ 1999, \mnras, 309, 803
\bibitem[Giveon et al.(1999)]{Giveon99} Giveon, U., Maoz, D., Kaspi, S., et al. \ 1999, \mnras, 306, 637
\bibitem[Hawkins(1993)]{Hawkins93} Hawkins, M. R. S. \ 1993, \nat, 366, 242
\bibitem[Hawkins(2000)]{Hawkins00} Hawkins, M. R. S. \ 2000, \aaps, 143, 465
\bibitem[Hook et al.(1994)]{Hook94} Hook, I. M., McMahon, R. G., Boyle, B. J., \& Irwin, M. J. \ 1994, \mnras, 268, 305
\bibitem[Ivezi\'c et al.(2004)]{ivezic04} Ivezi\'c, \v{Z}., Lupton, R., Johnston, D., et al. \ 2004a, in ASP Conf. Ser. 311, AGN Physics with the Sloan Digital Sky Survey, ed. G. T. Richards \& P. B. Hall (San Francisco, CA: ASP), 437
\bibitem[Ivezi\'c et al.(2007)]{ivezic07} Ivezi\'c, \v{Z}., Smith, J. A., Miknaitis, G., et al. \ 2007, \aj, 134, 973
\bibitem[Kasliwal et al.(2015)]{Kasliwal15} Kasliwal, V. P., Vogeley, M. S., \& Richards, G. T. \ 2015, \mnras, 451, 4328
\bibitem[Kato et al.(1996)]{Kato96} Kato, S., Abramowicz, M. A., \& Chen, X. \ 1996, \pasj, 48, 67
\bibitem[Kawaguchi et al.(1998)]{Kawaguchi98} Kawaguchi, T., Mineshige, S., Umemura, M., \& Turner, E. L. \ 1998, \apj, 504, 671
\bibitem[Kelly et al.(2009)]{Kelly09} Kelly, B. C., Bechtold, J., \& Siemiginowska, A. \ 2009, \apj, 698, 895 1, 1
\bibitem[Koz\l{}owski et al.(2010)]{Kozlowski10} Koz\l{}owski, S., Kochanek, C. S., Udalski, A., et al. \ 2010, \apj, 708, 927
\bibitem[Koz\l{}owski(2016)]{Kozlowski16r} Koz\l{}owski, S. \ 2016, \apj, 826, 118
\bibitem[Koz\l{}owski(2017a)]{Kozlowski16} Koz\l{}owski, S. \ 2017a, \apjs, 228, 9
\bibitem[Koz\l{}owski(2017b)]{Kozlowski17} Koz\l{}owski, S. \ 2017b, \apj, 835, 250
\bibitem[Lang et al.(2016)]{Lang16} Lang, D., David, W., \& Mykytyn, D \ 2016, ascl.soft 04008
\bibitem[Lewis \& Irwin(1996)]{Lewis96} Lewis, G. F., \& Irwin, M. J. \ 1996, \mnras, 283, 225
\bibitem[Li \& Cao(2008)]{LiCao08} Li, S.-L., \& Cao, X. \ 2008, \mnras, 387, L41
\bibitem[Lloyd(1984)]{Lloyd84} Lloyd, C. \ 1984, \mnras, 209, 697
\bibitem[Lupton et al.(1999)]{Lupton99} Lupton, R.H., Gunn, J.E., \& Szalay, A.S. \ 1999, \aj, 118, 1406
\bibitem[MacLeod et al.(2010)]{Mac10} MacLeod, C. L., Ivezi\'c, \v{Z}., Kochanek, C. S., et al. \ 2010, \apj, 721, 1014
\bibitem[MacLeod et al.(2012)]{Mac12} MacLeod, C. L., Ivezi\'c, \v{Z}., Sesar, B., et al. \ 2012, \apj, 753, 106
\bibitem[Matthews \& Sandage(1963)]{Matthews63} Matthews, T. A., \& Sandage, A. R. \ 1963, \apj, 138, 30
\bibitem[McHardy et al.(2006)]{McHardy06} McHardy, I. M., Koerding, E., Knigge, C., Uttley, P., \& Fender, R. P. \ 2006, \nat, 444, 730
\bibitem[Morganson et al.(2014)]{Morganson14} Morganson, E., Burgett, W. S.,  Chambers, K. C.. et al. \ 2014 \apj, 784, 92
\bibitem[O'Brien, Gondhalekar \& Wilson(1988)]{OBrien88} O'Brien, P. T., Gondhalekar, P. M., \& Wilson, R. \ 1988, \mnras, 233, 845
\bibitem[Oke \& Gunn(1983)]{Oke83} Oke, J. B., \& Gunn, J. E. \ 1983, \apj, 266, 713
\bibitem[Paltani \& Courvoisier(1994)]{Paltani94} Paltani, S., \& Courvoisier, T. J.-L. \ 1994, \aap, 291, 74
\bibitem[Paltani \& Courvoisier(1997)]{Paltani97} Paltani, S., \& Courvoisier, T. J.-L. \ 1997, \aap, 323, 717
\bibitem[P\^{a}ris et al.(2017)]{Paris17} P\^{a}ris, I., Petitjean, P., Ross, N. P., et al. \ 2017, \aap, 597, 79
\bibitem[Pica \& Smith(1983)]{Pica83} Pica, A. J., \& Smith, A. G. \ 1983, \apj, 272, 11
\bibitem[Rumbaugh et al.(2018)]{Rumbaugh18} Rumbaugh, N., Shen, Y., Morganson, E., et al. \ 2018, \apj, 854, 160
\bibitem[Rees(1984)]{Rees84} Rees, M. J. \ 1984, \araa, 22, 471
\bibitem[Rengstorf et al.(2006)]{Rengstorf06} Rengstorf, A. W., Brunner, R. J., \& Wilhite, B. C. \ 2006, \aj, 131, 1923
\bibitem[Schmidt et al.(2010)]{Schmidt10} Schmidt, K. B., Marshall, P. J., Rix, H.-W., et al. \ 2010, \apj, 714, 1194
\bibitem[Schneider et al.(2010)]{Schneider10} Schneider, D. P., Richards, G. T., Hall, P. B., et al. \ 2010, \aj, 139, 2360
\bibitem[Sesar et al.(2006)]{Sesar06} Sesar, B., Svilkovi\'c, D., Ivezi\'c, \v{Z}., et al. \ 2006, \aj, 131, 2801
\bibitem[Sesar et al.(2007)]{Sesar07} Sesar, B., Ivezi\'c, \v{Z}., Lupton, R. H., et al. \ 2007, \aj, 134, 2236
\bibitem[Shakura \& Sunyaev(1976)]{Shakura76} Shakura, N. I., \& Sunyaev, R. N. \ 1976, \mnras, 175, 613
\bibitem[Shen et al.(2011)]{Shen11} Shen, Y., Richards, G. T., Strauss, M. A., et al. \ 2011, \apjs, 194, 45
\bibitem[Smith \& Nair(1995)]{Smith95} Smith, A. G., \& Nair, A. D. \ 1995, \pasp, 107, 863
\bibitem[Terlevich et al.(1992)]{Terlevich92} Terlevich, R., Tenorio-Tagle, G., Franco, J., \& Melnick, J. \ 1992, \mnras, 255, 713
\bibitem[Torricelli-Ciamponi et al.(2000)]{Tor00} Torricelli-Ciamponi, G., Foellmi, C., Courvoisier, T. J.-L., \& Paltani, S. \ 2000, \aap, 358, 57
\bibitem[Tr\`evese et al.(1994)]{Trevese94} Tr\`evese, D., Kron, R. G., Majewski, S. R., Bershady, M. A., \& Koo,D.C. \ 1994, \apj, 433, 494
\bibitem[Uomoto, Wills \& Wills(1976)]{Uomoto76} Uomoto, A. K., Wills, B. J., \& Wills, D. \ 1976, \aj, 81, 905
\bibitem[van den Bergh et al.(1973)]{van73} van den Bergh, S., Herbst, E., \& Pritchet, C. \ 1973, \aj, 78, 375
\bibitem[Vanden Berk et al.(2004)]{Vanden Berk04} Vanden Berk, D. E., Wilhite, B. C., Kron, R. G., et al. \ 2004, \apj, 601, 692
\bibitem[Vanden Berk et al.(2001)]{Vanden Berk01} Vanden Berk, D. E., Richards, G. T., Bauer, A., et al. \ 2001, \aj, 122, 549
\bibitem[Webb \& Malkan(2000)]{Webb00} Webb, W., \& Malkan, M. \ 2000, \apj, 540, 652
\bibitem[Wilhite et al.(2008)]{Wilhite08} Wilhite, B. C., Brunner, R. J., Grier, C. J., Schneider, \& D. P., Vanden Berk. D. E., \ 2008, \mnras, 383, 1232
\bibitem[Willott et al.(2003)]{Willott03} Willott, C. J., McLure, R. J., \& Jarvis, M. J. \ 2003, \apj, 587L, 15
\bibitem[Wold et al.(2007)]{Wold07} Wold, M., Brotherton, M. S., \& Shang, Z. \ 2007, \mnras, 375, 989 1
\bibitem[Zackrisson et al.(2003)]{Zackrisson03} Zackrisson, E., Bergvall, N., Marquart, T., \& Helbig, P. \ 2003, \aap, 408, 17
\bibitem[Zuo et al.(2012)]{Zuo12} Zuo, W., Wu, X.-B., Liu, Y.-Q., \& Jiao, C.-L. \ 2012, \apj, 758, 104 1, 4.2


\end{thebibliography}
\end{document}